# Title: A transformer-based deep learning approach for classifying brain metastases into primary organ sites using clinical whole brain MRI

**Authors:** Qing Lyu[1], Sanjeev V. Namjoshi[2-3], Emory McTyre[4-6], Umit Topaloglu[2-3], Richard Barcus[5,7], Michael D. Chan[2,4,6], Christina K. Cramer[2,4,6], Waldemar Debinski[2-4], Metin N. Gurcan[2,9], Glenn J. Lesser[2,4,9], Hui-Kuan Lin[2-3], Reginald F. Munden[2,7], Boris C. Pasche[2-4], Kiran K. Sai[2,4-5,7], Roy E. Strowd[2,4,9,10], Stephen B. Tatter[2,4,6,8], Kounosuke Watabe[2,3], Wei Zhang[2,3], Ge Wang[1*], Christopher T. Whitlow[2,4-8,10-11*]

**Affiliations:**

[1]Department of Biomedical Engineering, Rensselaer Polytechnic Institute, Troy, NY, USA

[2]Comprehensive Cancer Center, Wake Forest School of Medicine, Winston-Salem, NC, USA

[3]Department of Cancer Biology, Wake Forest School of Medicine, Winston-Salem, NC, USA

[4]Brain Tumor Center of Excellence, Wake Forest School of Medicine, Winston-Salem, NC, USA

[5]Radiology Informatics & Image Processing Laboratory, Wake Forest School of Medicine, Winston-Salem, NC, USA

[6]Department of Radiation Oncology, Wake Forest School of Medicine, Winston-Salem, NC, USA

[7]Department of Radiology, Wake Forest School of Medicine, Winston-Salem, NC, USA

[8]Department of Neurosurgery, Wake Forest School of Medicine, Winston-Salem, NC, USA

[9]Department of Internal Medicine, Wake Forest School of Medicine, Winston-Salem, NC, USA

[10]Department of Neurology, Wake Forest School of Medicine, Winston-Salem, NC, USA

[11]Department of Biomedical Engineering, Wake Forest School of Medicine, Winston-Salem, NC, USA

*Corresponding author. Ge Wang (wangg6@rpi.edu), Christopher T. Whitlow (cwhitlow@wakehealth.edu).

**One Sentence Summary:** A transformer-based deep neural network accurately identifies the primary organ site of brain metastasis from a whole-brain MRI scan.

**Abstract:** Treatment decisions for brain metastatic disease rely on knowledge of the primary organ site, and currently made with biopsy and histology. Here we develop a novel deep learning approach for accurate non-invasive digital histology with whole-brain MRI data. Our IRB-approved single-site retrospective study was comprised of patients (n=1,399) referred for MRI treatment-planning and gamma knife radiosurgery over 21 years. Contrast-enhanced T1-weighted and T2-weighted Fluid-Attenuated Inversion Recovery brain MRI exams (n=1,582) were preprocessed and input to the proposed deep learning workflow for tumor segmentation, modality transfer, and primary site classification into one of five classes. Ten-fold cross-validation generated overall AUC of 0.878 [95%CI:0.873,0.883], "*Lung*" class AUC of 0.889 [95%CI:0.883,0.895], "*Breast*" class AUC of 0.873 [95%CI:0.860,0.886], "*Melanoma*" class

AUC of 0.852 [95%CI:0.842,0.862], "*Renal*" class AUC of 0.830 [95%CI:0.809,0.851], and "*Other*" class AUC of 0.822 [95%CI:0.805,0.839]. These data establish that whole-brain imaging features are discriminative to allow accurate diagnosis of the primary organ site of malignancy. Our end-to-end deep radiomic approach has great potential for classifying metastatic tumor types from whole-brain MRI images. Further refinement may offer an invaluable clinical tool to expedite primary cancer site identification for precision treatment and improved outcomes.

**Main Text:**

**INTRODUCTION**

Approximately 180,000 patients are diagnosed with brain metastases in the US each year (*1*). These metastases originate from primary organ site cancers across the body, with 67%-80% originating from lung, breast, or melanoma (*2, 3*). Clinical trials routinely assign treatment to patients with brain metastases based on the number and size of their brain lesions (*4*), often treating all lesions as equivalent diseases (*5*). It is becoming increasingly recognized that brain metastases of different histologies have correspondingly different biologic behaviors and therapeutic responses. For example, it has been known for many years that melanoma brain metastases are more likely to be hemorrhagic. Breast, lung and melanoma primaries are more likely to yield multiple brain metastases, while tumors of the gastrointestinal tract are more likely to seed the posterior fossa (*6*). Such patterns are readily visible on MRI examinations and have been previously noted in population studies. However, the overall sensitivity and specificity as applied to individual patients in the context of qualitative imaging evaluation of brain metastases is insufficient for definitive diagnosis/classification of the primary underlying malignancy. That said, there are also likely biological differences between brain metastases of various histologic types that have yet to be discovered and are not evident to physicians via standard qualitative approaches. This is currently reflected within the highly granular quantitative medical imaging data, which can be leveraged for sensitive and specific classification into brain metastatic disease histologic types. Our overall hypothesis is that imaging features of brain metastases are sufficiently discriminative and may be detected by radiomics at the time of initial evaluation to allow accurate diagnosis of the primary organ site cancer.

In approximately 10% of cases, brain metastatic disease is the initial cancer presentation (*7*), and treatment decisions are driven by knowledge of the primary organ site/histologic type. In these cases, an oncologic workup including cross-sectional imaging of the remainder of the body may identify the most likely primary organ site. If a primary malignancy site is identified that is amenable to biopsy, then tissue is often obtained directly from that extracranial site to confirm the histologic diagnosis. If there are no extra-cranial sites amenable to biopsy, and the brain metastatic disease is in a favorable location, then invasive intracranial tissue sampling may be necessary for definitive diagnosis. Often, the specific treatment recommended for brain metastatic disease hinges on knowing the primary organ site of the malignancy, which underscores the importance and immediate need to develop a non-invasive methodology that can conclusively identify the primary organ site of brain metastatic disease to facilitate rapid management and optimize treatment decision.

Deep learning-based radiomics, as an emerging field, uses deep learning-based algorithms to transfer standard medical images into high-dimensional quantitative features, have led to new tools capable of recognizing features in medical imaging data that may not be easily perceived by physicians, such as radiologists (*8*). Radiomic tools have been applied to quantitative image analysis and expedite the clinical diagnosis of cancer (*8-10*). To date, physicians who use diagnostic imaging for neuro-oncology evaluation, such as radiologists and oncologists, are generally unable to determine the primary organ site of cancer from images of brain metastases alone. However, it is possible to use radiomic tools to identify the primary organ site of brain metastases by leveraging patterns or features within medical imaging data (*11*). Previously, radiomic tools utilized dedicated algorithms to extract handcrafted features. As a result, only a limited number of heuristic features are available. Although such a problem can be alleviated with the image transform like wavelet transform (*11*), the scope of extracted features is still limited. Deep learning, as the mainstream of artificial intelligence, is a data-driven approach, holding a great promise to address this challenge. Recently, deep learning has been widely used in radiomics for many tasks, such as cancer prognostication and cancer radiotherapy failure rate prediction (*12, 13*). Recently, deep learning has also achieved great success in recognizing primary or metastasis tumors and classifying metastatic tumors into their origins based on whole-slide histological images (*14*).

We hypothesized that deep learning-based radiomics, or deep radiomics, could identify the primary organ site cancer associated with a patient's brain metastatic disease using clinical MR images. Furthermore, we hypothesized that the classification process could be expedited with minimal image preprocessing in an automated workflow that requires minimal human intervention/feature engineering. Here we report a deep radiomic approach for classification of the primary organ site of brain metastases based on 3D structural MRI T1 and/or T2-weighted-Fluid-Attenuated Inversion Recovery (T2 FLAIR) images.

## RESULTS

### Patient demographic analysis

Two large datasets with a total of 1,399 patients (mean ± SD age, 62.77 ± 11.86 years, 63.57% male) were used. Specifically, for the dataset with tumor contours, there were 148 patients (61.77 ± 11.15 years; 56.85% male); and for the dataset without tumor contours, there were 1,251 patients (62.85 ± 11.91 years; 59.07% male). Specifically, the "*Lung*" category had 801 patients (63.21 ± 11.06 years; 54.93% male), "*Breast*" category had 106 patients (62.23 ± 12.04 years), "*Melanoma*" category had 313 patients (62.80 ± 13.81 years; 72.20% male), "*Renal*" category had 82 patients (62.47 ± 11.16 years; 85.37% male), and "*Other*" category had 97 patients (59.81 ± 12.25 years; 82.47% male). In the dataset with tumor contours, 54.11% patients/scans had 1-3 lesions (mean ± SD = 4.79 ± 5.27).

### Tumor segmentation

To conduct deep radiomic analysis, it is necessary to label targets and extract their features. In this study, we adapted a U-Net-shaped network with transformers in the bottleneck for tumor segmentation. The proposed segmentation network segments brain metastasis tumors and generates voxel-wise tumor probability maps to guide the proposed classification network to pay due attention on metastases. The proposed networks were trained on T1 CE and T2 FLAIR dataset separately. We compared our results with that from some popular medical segmentation networks including U-Net (*15*), Attention U-Net (AttU-Net) (*16*), and U-Net Plus Plus (U-Net++) (*17*). Tumor segmentations were evaluated both qualitatively and quantitatively.

Figure 3 compares our proposed method with other brain tumor segmentation networks. Different from the other methods, our proposed network has two major merits: 1) high sensitivity on both smaller and larger metastases, 2) high robustness over different datasets. For the brain metastasis segmentation, one major challenge is a strong heterogeneity of tumor sizes. Some small metastasis tumors only account for a few voxels in an MR image, which can be easily neglected by generic networks since information can be lost during a down-sampling process. On the other hand, how to precisely segment a large tumor is also difficult because convolutional neural networks (CNN) tend to focus more on local features. To address this problem, we proposed a novel network synergizing both convolutional layers and transformers. In contrast to convolutional layers, transformers focus on global information via a self-attention mechanism. The combination of locally oriented convolutional layers with transformers with global awareness guarantees that the proposed network integrates both local and contextual features. In additional, the skip connections are made between down-sampling and up-sampling blocks to minimize the information loss in the feature extraction process. By this architectural design, our proposed network has inherent capabilities of segmenting both small (e.g., the tumor pointed by the yellow arrow) and large metastasis tumors (e.g., the tumor pointed by the red arrow) equally well. Interestingly, in some cases, segmented results could outperform human labeled results, as the tumor pointed by the green arrow.

The other major merit of the proposed segmentation network is its robustness and generalizability over different datasets. When testing on data without tumor contours (Figure 3b), the proposed network behaved quite "conservatively" to avoid false positive prediction. As shown in the third row in Figure 3b, the proposed network only segmented the region with strong tumor feature while the other methods can easily generate false positive prediction. Also, as indicated by the yellow arrow, our method can segment tiny tumors on unlabeled data. Because a majority of data used in this study contains no tumor contour, it is difficult to directly evaluate the segmentation accuracy on the unlabeled dataset. To ensure the proposed network effective on unlabeled data, we prefer that the proposed network can be more "conservative" when reporting a region as a tumor than these other methods. Such a design would effectively reduce the false positive rate and improve the robustness on unlabeled data. Network's robustness is of critical importance in the clinical setting. As MRI images are collected from scanners manufactured by different manufactures, it is impractical to have every MRI case collected under the same protocol and techniques. A robust network can tolerate the conditional differences in the data collection process and generalize well on different datasets.

Quantitative comparison of the segmentation results from different methods are shown in Table 3. It can be found that the results obtained using the proposed method have higher Dice score than the other methods on both T1 CE and T2 FLAIR datasets, showing the effectiveness of the proposed segmentation network.

**Modality transfer**

In this study, we used both of T1 CE and T2 FLAIR images for tumor classification so that complementary features could enhance the radiomic analysis. We compared our modality transferred results with another method called style transfer (*18*). Modality transfer results were presented in Figure 5. It was found that cycleGAN can generate modality transferred results much closer to the target modality than those from the style transfer method. We tested cycleGAN on each patient only with either T1 CE or T2 FLAIR images to generate the missing modality image, which were used so that our classification network always had T1 CE and T2 FLAIR images as the input.

**Five-class classification**

Using the proposed tumor segmentation network and cycleGAN, we obtained a set of data for each MRI scan: the original MRI scan in the either T1 CE or T2 FLAIR contrast mechanism, its corresponding missing modality data, and voxel-wise tumor probability map. We fed these images and map to the classification network for predicting the primary organ site of brain metastases. Table 4 show our 10-fold classification results. The overall AUC is 0.889 with 95% confidence interval of 0.873 and 0.883; "*Lung*" class AUC is 0.889 with 95% confidence interval of 0.883 and 0.895; "*Breast*" class AUC of 0.873 with 95% confidence interval of 0.860 and 0.886, "*Melanoma*" class AUC is 0.852 with 95% confidence interval of 0.842 and 0.862; "*Renal*" class AUC is 0.830 with 95% confidence interval of 0.809 and 0.851; and "*Other*" class AUC is 0.822 with 95% confidence interval of 0.805 and 0.839. Figure 7c shows the top-k classification results. Compared with the top-1 accuracy, top-2 and top-3 results are with a higher accuracy in each category. Top-3 achieved the highest prediction accuracy, and the differences between different categories are smaller than that for top-1 and top-2.

**Ablation study**

To investigate the contribution of adding tumor prediction maps to the performance of the classification network, we conducted an ablation study without involving tumor prediction maps. It can be observed in Figure 7d and Table 5 that adding tumor classification maps can improve the performance of the proposed network with an overall AUC increasing by 2.3%. Meanwhile, another ablation study was performed only using T1 CE or T2 FLAIR images for tumor classification. Based on our results, the removal of either T1 CE or T2 FLAIR images and only used single modality images for classification decreased the performance of the proposed network. Only using T1 CE or T2 FLAIR images will reduce the overall AUC score by 3.8% and 3.1%, respectively.

**Binary classification**

Apart from the five-class classification, we also conducted a series of experiments on binary classification. Due to the limited number of cases in the "*Renal*" and "*Other*" categories, we combined these two categories together and labeled as a new "*Other*" category. Then, the data in four different classes were used for binary classification. We trained multiple classification networks separately, in the same setting as the five-class classification experiment except changing the weights in the weighted binary-cross-entropy and the sampling strategy in the training process. During the training, 60% samples were randomly selected from the category in each epoch.

We summarized all binary classification results in Table 6. Compared with the five-class classification results, all binary classification results are with higher AUC scores, indicating better classification performance. Among all binary classification results, the proposed network achieved the best performance when distinguishing the "Lung" and "Breast" categories, yielding an AUC score of 0.959.

**DISCUSSION**

Brain metastases are common and often present as the initial presentation of lung, melanoma, renal, breast, and other cancers (*19, 20*). Of those who develop brain metastases, 36% are diagnosed within 1-month of the initial presentation and 72% are diagnosed within the first year (*21*). Survival of patients presenting with brain metastatic disease is poor, with a median survival of 3.7 months and worse particularly for patients with lung cancer (*22*). Factors associated with more favorable survival for these patients with brain metastatic disease include use of appropriate systemic therapy for better performance status (*23*). Critical to initiation of optimal systemic therapy is establishing an accurate diagnosis of the primary organ site. Of particular importance is not delaying initiation of brain-directed treatment with extensive systemic imaging studies, diagnostic biopsies, or other methods for primary organ site identification (*24-26*). To this end, our work presents a deep learning-based radiomic pipeline for easy and conclusive identification of the most common primary organ sites for brain metastatic disease using clinically acquired T1 CE or T2 FLAIR whole brain MRI images alone, thus opening the door for rapid translation into clinical care. Owing to the end-to-end nature, the proposed framework is user friendly and can be implemented within seconds with minimal human intervention. To our best knowledge, this study represents the first application of deep learning to extract radiomic features for this critically important classification task, and the cross-validation approach suggests that the results are generalizable.

Our results further demonstrate that the deep learning approach is feasible and could be improved with a larger image dataset. We achieved the highest accuracy on the "*Lung*" category, which incorporated the majority (55.82%) of all cases used in this study. Since deep learning is data-driven, the network can better learn how to extract radiomic features for analysis with more cases involved in the training process. Within one month after the publication of this paper, we will open our codes for federated learning, and continuously add more whole-brain MRI cases into the proposed radiomic pipeline to further improve the classification performance.

The inclusion of cycleGAN in the proposed radiomic pipeline relaxes the requirement for clinical translation. Clinically, it is rare to have a large batch of datasets with each subject scanned in multiple MRI modalities. A common scenario is that an individual patient is scanned with only a subset of MR imaging sequences. The use of cycleGAN for modality transfer generates missing modalities semantically, overcoming the difficulty of data collection and making the proposed method widely applicable. Meanwhile, more single modality cases can be capitalized via modality transfer in the training process to improve classification results. As shown in Figure 7d and Table 5, compared with using either T1 CE or T2 FLAIR alone, the dual-modality input via modality transfer led to significantly higher AUC scores.

There are several opportunities for further improvements of our deep learning model. First, additional medical images from the primary cancer site should be helpful in the proposed radiomic pipeline, such as lung CT images. We hypothesize that there may be some common or correlated deep radiomic features between tumors in the primary site and the brain. Adding these additional images may guide the network to find these common or correlated features for better classification. While MRI is uncommon in certain body regions, there are already efforts underway to supplement patient screening in domains currently dominated by radiography and

CT (*27*). Second, with a greater number of images, it will be possible to increase the number of classes and cover more histological subtypes. Brain metastases originating from different subtypes may have differential responses to certain targeted treatments and immunotherapies, and a finer-grained classification scheme may be necessary to elucidate these differences. For example, Lu *et. al.* divided metastasis into 18 categories based on 32,537 whole-slide images (*14*). Third, to obtain more data for deep learning, the partnership or consortium is important among multiple institutions and hospitals. To address the concern over privacy and ownership of medical data, one promising solution is federated learning (*28*). Fourth, further iterations of our model will be needed to diagnose patients with cancer of unknown primary, which may remain unknown until death in some cases. Hopefully, we will refine our model in the future to further analyze patients with cancer of unknown primary for the true etiology so that optimal therapies can be initiated in a timely manner. We had initially sought to test our model on cancer of unknown primary; however, there were too few cases in our cohort (n < 21).

Using deep learning to extract radiomic features has shown a great potential to outperform manually driven radiomic analyses (*11, 29*). For this brain metastasis classification, beyond the previous manually driven feature-derived results, our approach has produced competitive AUC scores in performing three-class classification (0.878 vs 0.873). Given the fact that our five-class classification is much harder than the mentioned three-class classification, our deep learning-based method showed advantages over those manually driven feature-derived methods. Unlike conventional radiomics relying on limited manually generated features like gray-level co-occurrence and run-length metrics, deep radiomics does not rely on pre-defined algorithms, instead it learns how to extract radiomic features by itself. For conventional radiomics, one typical scenario is that if manually derived features are not adequately extracted, radiomics cannot obtain good results. Such a problem can be solved by deep radiomics through data-driven learning. Although it is in principle that deep learning-based radiomics is superior to conventional methods, there is no need to discard conventional radiomics. A synergistic way is to combine conventional manually derived feature extraction with deep radiomics. While deep radiomics is automatic and systematic, manually derived features have clearer meanings and rich semantics. In other words, learned and handcrafted features may help each other to boost the diagnostic performance.

Last but not the least, deep learning-mediated classification of the primary organ site cancer associated with brain metastatic disease could improve outcomes in the setting of healthcare disparities. According to a recent CDC study, 263,054 adult respondents to a survey in the United States came from rural counties (*30*). These counties generally have less access to healthcare, with hospitals that may be understaffed and lacking the equipment to perform specialized diagnostic tests. Such health disparities have resulted in poorer health status and quality of life, as compared to the general population. Indeed, Renz et al. (*31*) noted that rural location is a predictor for increased death, as well as compromised whole brain radiation therapy outcomes for squamous cell carcinoma brain metastases. An algorithm like the one described in this paper, once properly trained and validated, can be deployed online to process images in real-time as a tool for radiologists and oncologists, to improve workflow and diagnosis. This would allow lower cost, fewer tests, and faster determination of the primary organ site of brain metastases, as well as less healthcare disparities. Collectively, our results show the potential of the deep learning approach that will have a widespread impact on the diagnosis and treatment of brain metastases to improve healthcare outcomes and quality of life for patients.

## MATERIALS AND METHODS

Figure 1a shows an overview of our study design. Two large datasets were clinically collected, consisting of 1,399 patients with 1,582 cases respectively. The images with tumor contours were used for training and validation of our tumor segmentation network. Once trained, voxel-wise tumor probability maps were generated by the tumor segmentation network. Patients with both T1-weighted contrast enhanced (T1 CE) and T2 FLAIR images were used to train a cycleGAN (*32*) in the weakly supervised learning mode to transfer image modalities between T1 CE and T2 FLAIR. After training, either a T1 CE or T2 FLAIR image can be used to produce the image in the complementary modality (that is, either the corresponding T2 FLAIR or T1 CE image). Finally, we used all cases, including modality transferred case corresponding to each case, and corresponding tumor probability maps in our classification network for inferencing the primary organ site of brain metastases. Once trained, the whole workflow can be implemented in an end-to-end way.

### Data acquisition and processing

For this IRB-approved single-site retrospective patient study, our initial dataset consists of 1,862 conventional treatment-planning MRI scans from 1,650 patients imaged at Wake Forest School of Medicine from 2000 to 2021, who were referred for Gamma Knife Radiosurgery. This study used structural MR images with patients in a stereotactic head frame as part of the clinical GK treatment planning routine on the following scanners: 1.5T General Electric SIGNA Excite (GE Healthcare), 1.5T General Electric SIGNA HDxt (GE Healthcare), 3.0T General Electric SIGNA Excite (GE Healthcare), and 3.0T Siemens Skyra (Siemens Medical Solutions).

### Image labeling and pre-processing

Clinical labels to the MR images were extracted from the electronic medical records, including tissue biopsy proven pathologic diagnoses. In total, there are 77 diagnostic categories for all the images. Apart from lung, melanoma, renal, and breast cancer, all other diagnostic labels were pooled together into an "*Other*" class. In other words, this labeling strategy defines five classes. All the images associated with a biopsy-proven diagnosis were included for network training.

Our images were collected in two batches: one batch of images had human labeled brain metastasis tumor contours, and the other batch of images were not processed and associated with no tumor contour. The original dataset with tumor contours contains 148 patients (Male: 83, Female: 65) with 360 cases (T1 CE: 171, T2 FLAIR: 189). In each case, tumor contours were manually labeled by experienced radiologists. Images were stored in the Dicom format, with the tumor contours stored in RTSTRUCT files. After the registration between the contours and the tumors, 77 cases were removed due to signal degradation, motion blurring, and other artifacts. In most of these failure cases, the main reason was the fading or diminishing of metastases after radiotherapy. There are 1,608 patients (Male: 981, Female: 627) in the dataset without tumor contours. Each patient was only scanned once. In this dataset, 182 cases were first excluded because they are neither T1 CE or T2 FLAIR. Then, additional 221 cases were removed because they failed to cover the whole brain or there were severe image quality issues. After exclusion of the inappropriate cases, 1,582 cases from 1,399 patients were selected for this study. Among them, 284 cases (T1 CE: 138, T2 FLAIR: 146) were with tumor labels. For each case, we unified voxel resolution and normalized intensity distributions. For the voxel resolution unification, we rescaled every whole-brain MRI scan to unify the voxel resolution as $0.64mm \times 0.64mm \times 2.4mm$.

**Training and validation datasets**

For tumor segmentation, we only utilized images with tumor contours for network training and validation. We randomly selected 120 T1 CE cases and 125 T2 FLAIR cases for training. The other 18 T1 CE cases and 20 T2 FLAIR cases were used for validation. Specifically, the proposed tumor segmentation network was trained on T1 CE and T2 FLAIR images respectively. After training, all cases without contour labels were processed by the network to generate the corresponding tumor segmentation probability maps using the Softmax function in the last layer of the network.

To perform the modality transfer between T1 CE and T2 FLAIR images, we used 254 cases (T1 CE: 127, T2 FLAIR: 127) from 127 patients to train and validate the modality transfer network. 230 cases from 115 patients were randomly selected for the training and the other 24 cases from the remaining 12 patients for validation. Once trained, all the cases used in this study were processed by the network to generate the modality transferred results.

For tumor classification, we utilized the 10-fold cross-validation scheme on all the cases used in this study involved in the training stage. The classification results shown in this paper are based on the average data of 10 cross-validation runs.

**Tumor segmentation network design**

The segmentation technique aims at extracting tumors from the whole-brain MRI scan to guide the classification network so that more attention can be paid to a tumor and its surrounding region for better classification results. In this study, we proposed an advanced network for brain tumor detection by combining convolutional layers for local feature extraction and transformers for global awareness. The proposed network is in the U-Net structure. As shown in many other deep learning tasks, adopting the U-Net demands a reasonable computational cost and delivers reasonable network performance. The structure of the proposed segmentation network is illustrated in Figure 2a-d.

The proposed network includes the four components: a down-sampling convolutional branch, a transformer-based bottleneck, an up-sampling convolutional branch, and skip connections between the down-sampling and up-sampling branches. The down-sampling branch contains ten down-sampling blocks. In each down-sampling block, there are two arms, the first arm has three convolutional layers follower by group normalization and ReLU activation. Down-sampling happens in the second convolutional layer with a stride of two. The second arm only has one convolutional layer with a stride of two. Feature maps from the two arms are finally added together. In the up-sampling branch, there are five up-sampling blocks. Each up-sampling block contains two convolutional layers followed by group normalization and ReLU activation. An up-sampling layer is then added to expand the feature maps by a factor of two. There are skip connections between each pair of down-sampling and up-sample blocks. The bottleneck consists of twelve identical transformers. Before feeding feature maps to the first transformer encoder, feature maps are vectorized into a series of 1D tokens. Combining convolutional layers and transformers ensures that the network would utilize both local and global information. For more details, please see Supplementary Table 1.

The objective function of the proposed segmentation network combines cross-entropy and Dice loss:

$$\min_{S} \mathcal{L} = 0.5 \; cross\_entropy + 0.5 \; Dice \qquad (1)$$

where $S$ is the parameters of the proposed segmentation network.

**Modality transfer network design**

For the data used in this study, there are only a small part of patients (127 out of 1,399) with both T1 CE and T2 FLAIR scans and most patients are only with either T1 CE or T2 FLAIR scans. Hoping to use both types of images for classification, we developed a method to generate missing images through converting an existing modality scan into the missing modality scan, as shown in Figure 4. CycleGAN (*32*) has shown great successes in image style transfer and other tasks. It uses a generative adversarial mechanism in the training process to let the generator learn the real distribution of target data. Meanwhile, the adoption of the cycle consistency loss avoids the contradiction in adversarially generated images, and the resultant cycleGAN network can be trained in a weakly supervised fashion. We used cycleGAN for modality transfer in this study. The proposed cycleGAN has two identical generators and two identical discriminators, as shown in Figure 3a. Generator 1 was designed to transform T1 CE images to T2 FLAIR counterparts, and Generator 2 was expected to generate T1 CE images from T2 FLAIR counterparts. Discriminators 1 and 2 judge the generated T1 CE or T2 FLAIR images are true or not. The generators and discriminators are of the same structures as those described in our recent work (*33*)

In the training process, mutual information is used to find the most similar slice from the corresponding modality scan of the same patient so that the network can be trained in a weakly supervised way. We use the least-square adversarial loss (*34*) in the training process. The objective function for the two generators is as follows:

$$\min_{G} \mathcal{L} = \mathcal{L}_{adv} + \lambda \mathcal{L}_{cyc} \qquad (2)$$

$$\mathcal{L}_{adv} = \mathbb{E}_x(D_1(G_1(x)) - 1)^2 + \mathbb{E}_y(D_2(G_2(y)) - 1)^2 \qquad (3)$$

$$\mathcal{L}_{cyc} = \mathbb{E}_x(G_2(G_1(x)) - x)^2 + \mathbb{E}_y(G_1(G_2(y)) - y)^2 \qquad (4)$$

where $G$ and $D$ stand for Generator and Discriminator respectively, $\mathcal{L}_{adv}$ is the adversarial loss, and $\mathcal{L}_{cyc}$ is the cycle consistency loss. The objective functions of the two generators are:

$$\min_{D_1} \mathcal{L} = \mathbb{E}_x(D_1(G_1(x)) - 0)^2 + \mathbb{E}_y(D_1(y) - 1)^2 \qquad (5)$$

$$\min_{D_2} \mathcal{L} = \mathbb{E}_y(D_2(G_2(y)) - 0)^2 + \mathbb{E}_x(D_2(x) - 1)^2. \qquad (6)$$

**Tumor classification network design**

As shown in Figure 6, the proposed tumor detection network consists of two identical feature extraction branches, where the probability maps are gradually down-sampled via 3D maximum pooling and finally concatenated with the features from the two branches. There is an attention module to combine the features extracted from T1 CE and T2 FLAIR branches. Then, two fully connected layers are used to generate the final classification outcome. Each feature extraction branch works through a 3D convolutional layer and five subsequent down-sampling blocks. The down-sampling blocks share the same structure as that in the segmentation network. The only difference lies in that 2D convolutional layers are upgraded to 3D convolutional layers. The attention module consists of a channel attention branch and a spatial attention branch. In the channel attention branch, feature maps are first converted into vectors via 3D global averaging, then two fully connected layers calculate the weight for each channel, and finally feature maps are channel-wise multiplied by channel weights. In the spatial attention branch, feature maps are first channel-wise averaged, the weight in each voxel is determined via sigmoid activation, and finally feature maps are voxel-wise multiplied by weights. The outputs of the channel attention and spatial attention branches are element-wise added to form new feature maps.

As shown in Table 1, the data used in this study for classification is highly unbalanced with the "*Lung*" category has far more cases than all the other categories, which is also in accordance with the well-known fact that the lung cancer is the major cause for all brain metastases cases. To overcome this unbalanced data problem, in the training process we used the weighted cross-entropy as the objective function for the proposed classification network. Also, we adopted over-sampling and under-sampling strategies during the training. In each epoch, 35% data input into the network was randomly selected from the "*Lung*" category. The contributions of data from "*Breast*", "*Melanoma*", "*Renal*", and "*Other*" categories were made 15%, 20%, 15% and 15% respectively,

**Implementation details**

We used the Adam optimizer to train the segmentation, modality transfer, and classification networks respectively. For the segmentation network, the batch size was set to 12 per GPU. The learning rate was $1e^{-4}$. The training stopped when there was no significant (less than 1%) loss decay for 20 epochs. For the modality transfer network, the batch size was 1 per GPU. The training continued 50 epochs with a learning rate of $1e^{-4}$. For the classification network, ten-fold cross-validation was adopted and results were obtained by averaging results from each fold. For each fold, we trained a classification network separately with the batch size of 3 per GPU. The learning rate was set to $1e^{-5}$ in the first 50 epochs and then divided by a factor of 2 after every 10 epochs. The training stopped when there was no significant loss decay for 20 epochs. All experiments were conducted on eight Nvidia Tesla V100 GPUs with 32 GB memory.

**Acknowledgments**

We would like to thank Nikita Namjoshi and Josh Tan for comments and suggestions. We would also like to thank Drs. Guangxu Jin and Liang Liu with Wake Forest Baptist Comprehensive Cancer Center Bioinformatics Shared Resource for their inputs.

**Funding:**
National Institutes of Health grant R01EB026646, R01CA233888, R01HL151561, R21CA264772, and R01EB031102 (QL, GW).
National Cancer Institute grant P01CA207206 and P30CA012197 (CW).
National Cancer Institute grant P01CA207206 and R01CA074145 (WD).

**Author contributions:**
Conceptualization: CW, GW, SN

Methodology: QL, SN
Investigation: CW RB, EM, UT
Visualization: QL, SN, GW, CW
Funding acquisition: CW, GW, WD
Project administration: CW, GW, WZ
Supervision: CW, GW, WZ
Writing – original draft: QL, SN, GW, CW
Writing – review & editing: All authors

**Competing interests:** The authors declare no competing interests.

**Data and materials availability:** All codes underlying this article will be available at [https://github.com/QingLyu0828/whole-brain-MRI_metastases_cls](https://github.com/QingLyu0828/whole-brain-MRI_metastases_cls). To access whole-brain MRI data used in this paper, please contact the corresponding author Christopher T. Whitlow. All data will be available after the approval of materials transfer agreements (MTAs).

# Figures and Tables

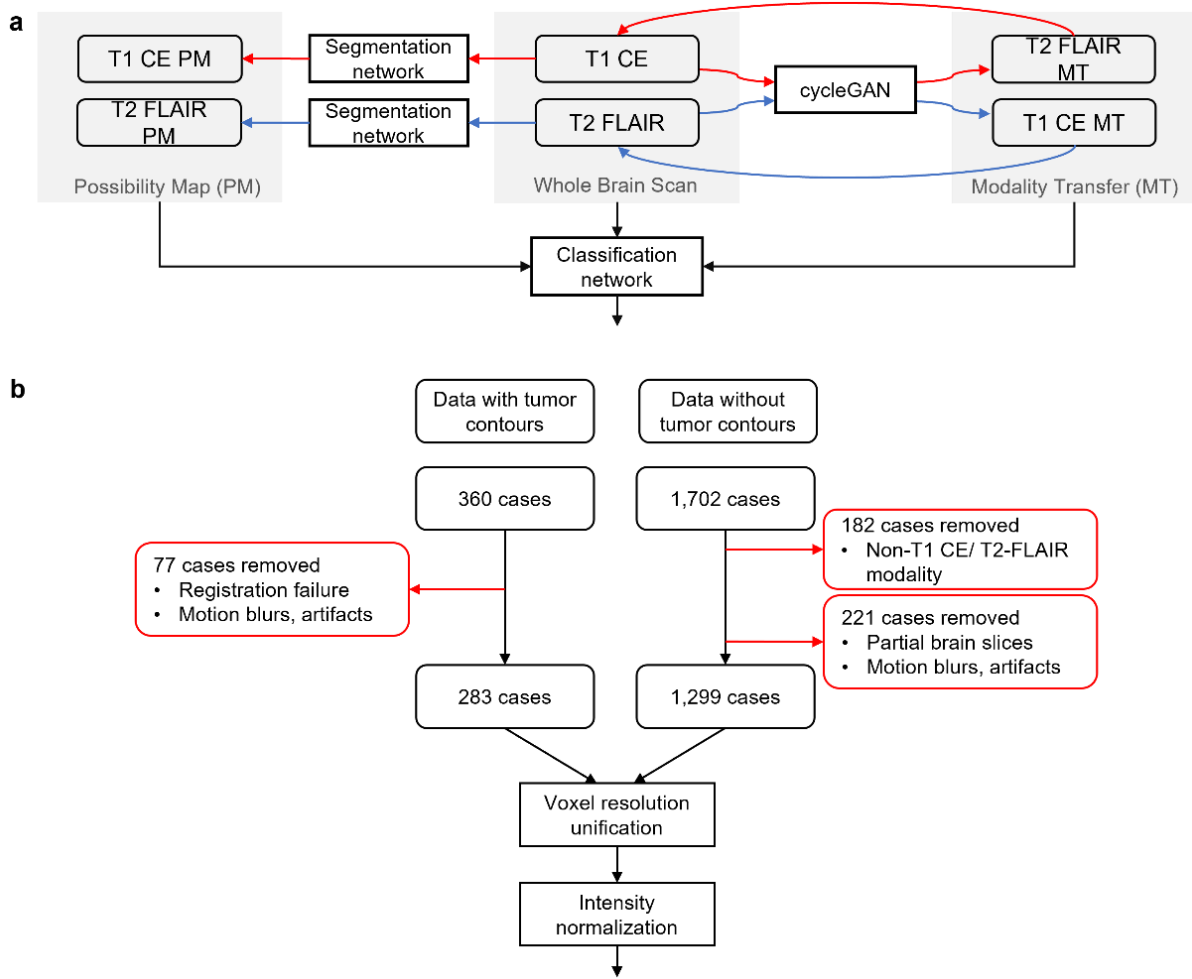

**Fig 1. Proposed deep radiomic workflow and data characteristics.** (a) Flowchart of data pre-processing. (b) Flowchart of the proposed deep learning-based pipeline for brain metastasis classification.

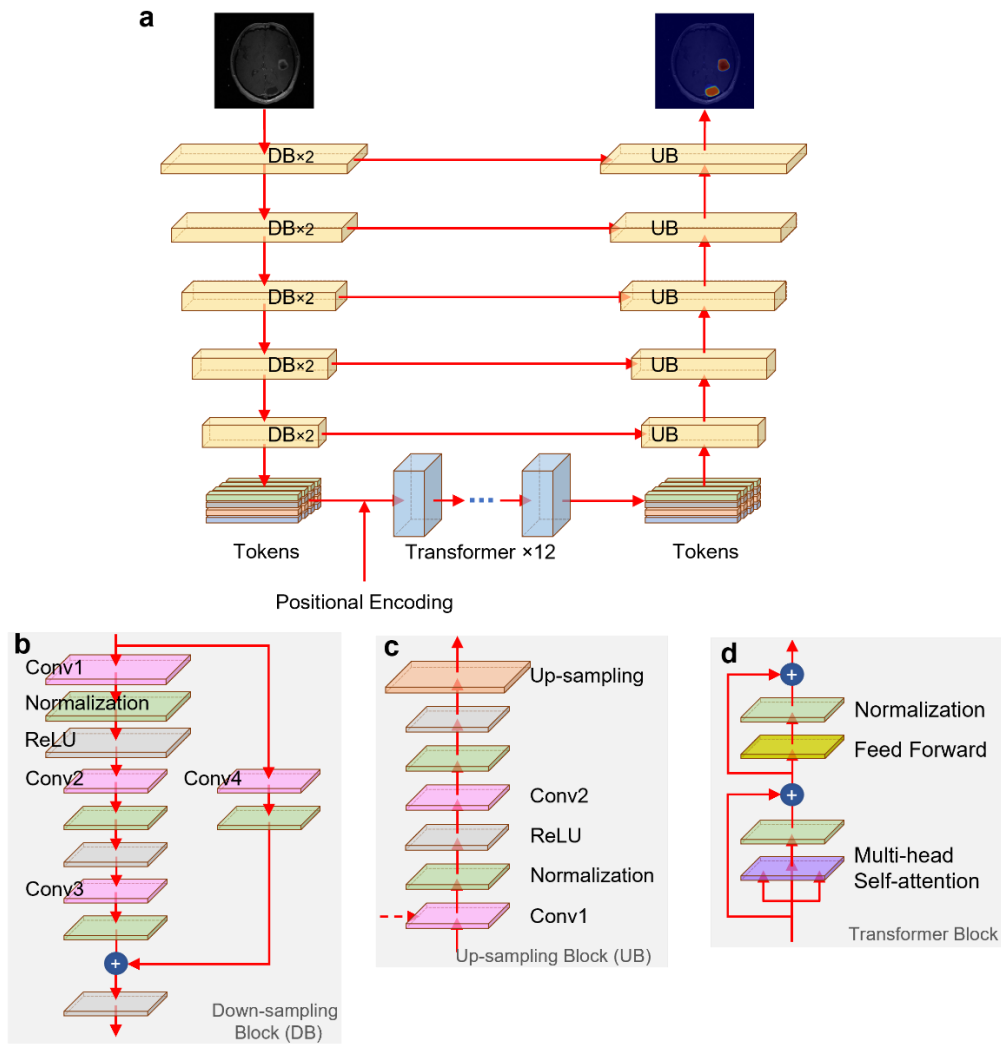

**Fig. 2. Proposed brain tumor segmentation network.** (a) The proposed segmentation network consisting of the four components: the down-sampling branch, up-sampling branch, transformer bottleneck, and skip connections for the U-net. (b-c) The structure of the down-sampling block and up-sampling block respectively. (d) The structure of transformer.

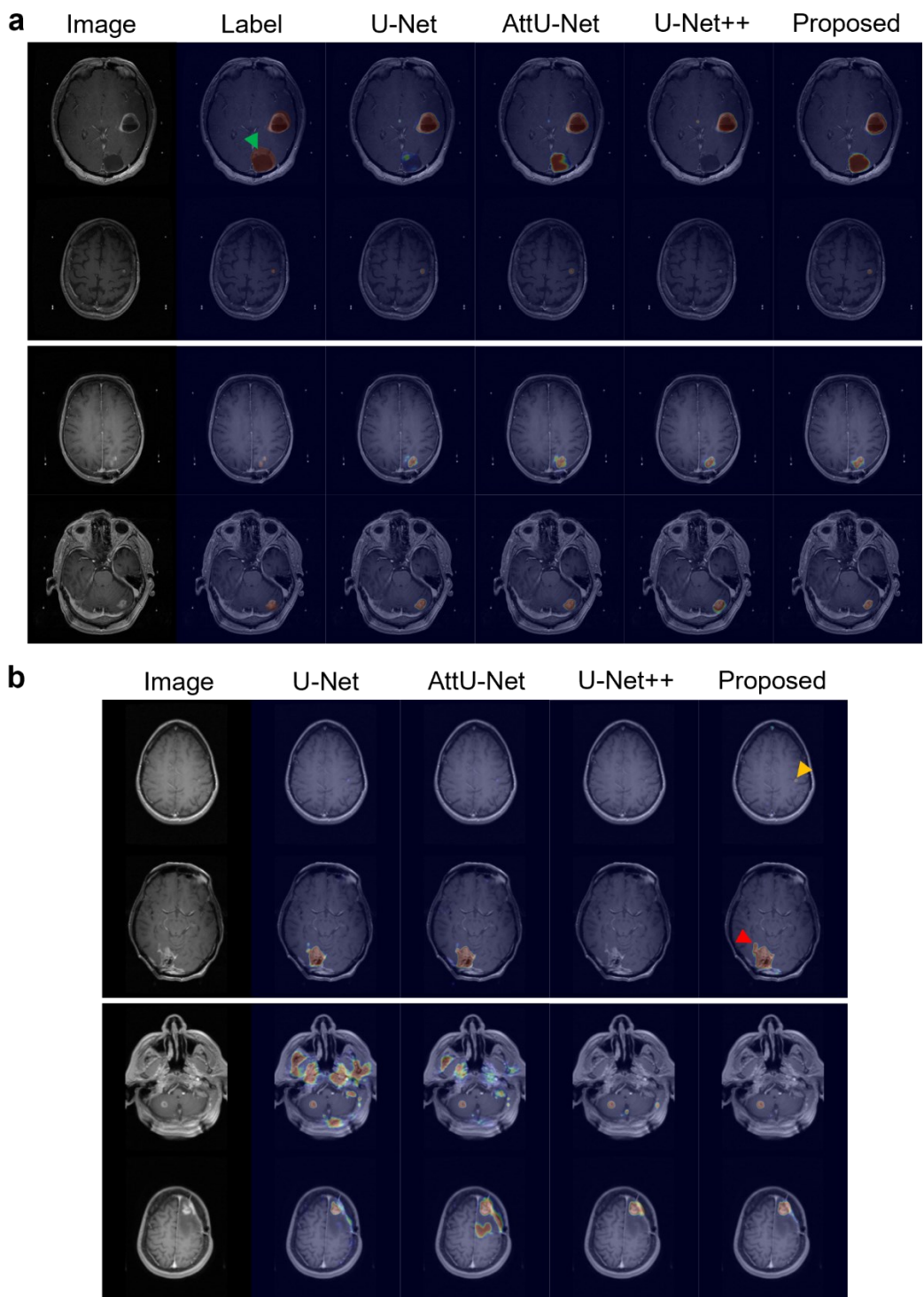

**Fig. 3. Comparison of tumor segmentation results.** (a) Segmentation results on the test dataset. (b) Segmentation results on the dataset without tumor contours. In each subfigure, the two top rows show T1 CE results, while the two bottom rows show T2 FLAIR results.

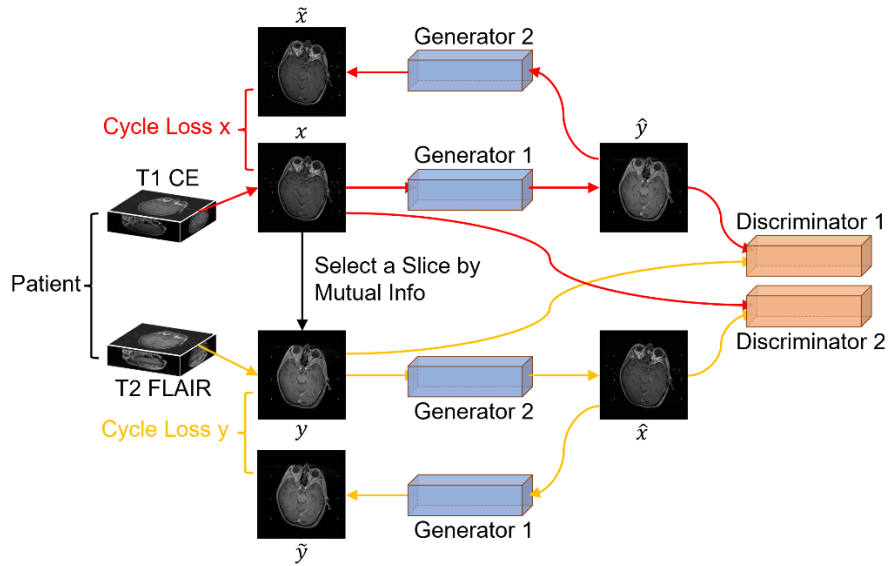

**Fig. 4. The schematic of the cycleGAN used for modality transfer**. There are two generators and two discriminators, where Generator 1 transforms T1 CE to T2 FLAIR, and Generator 2 generates T2 FLAIR from T1 CE. In the training process, mutual information is used to find the most similar slice from the corresponding modality scan of the same patient for weakly supervised learning.

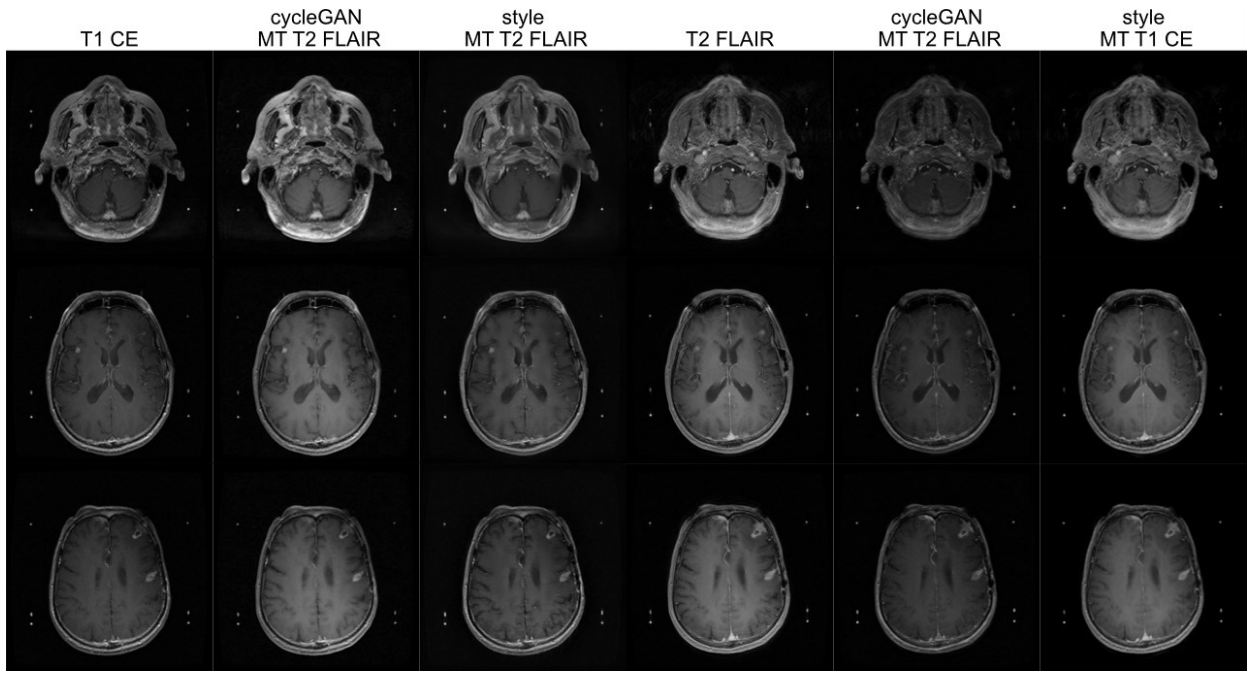

**Fig. 5. Comparison of original T1 CE and T2 FLAIR images and their modality transfer results**. The first column shows the original T1 CE images, the second and third columns give the modality transferred T2 FLAIR images from cycleGAN and style transfer methods respectively; the third column shows the original T2 FLAIR images, and the fourth and fifth columns are the modality transferred T1 CE images from cycleGAN and style transfer methods respectively.

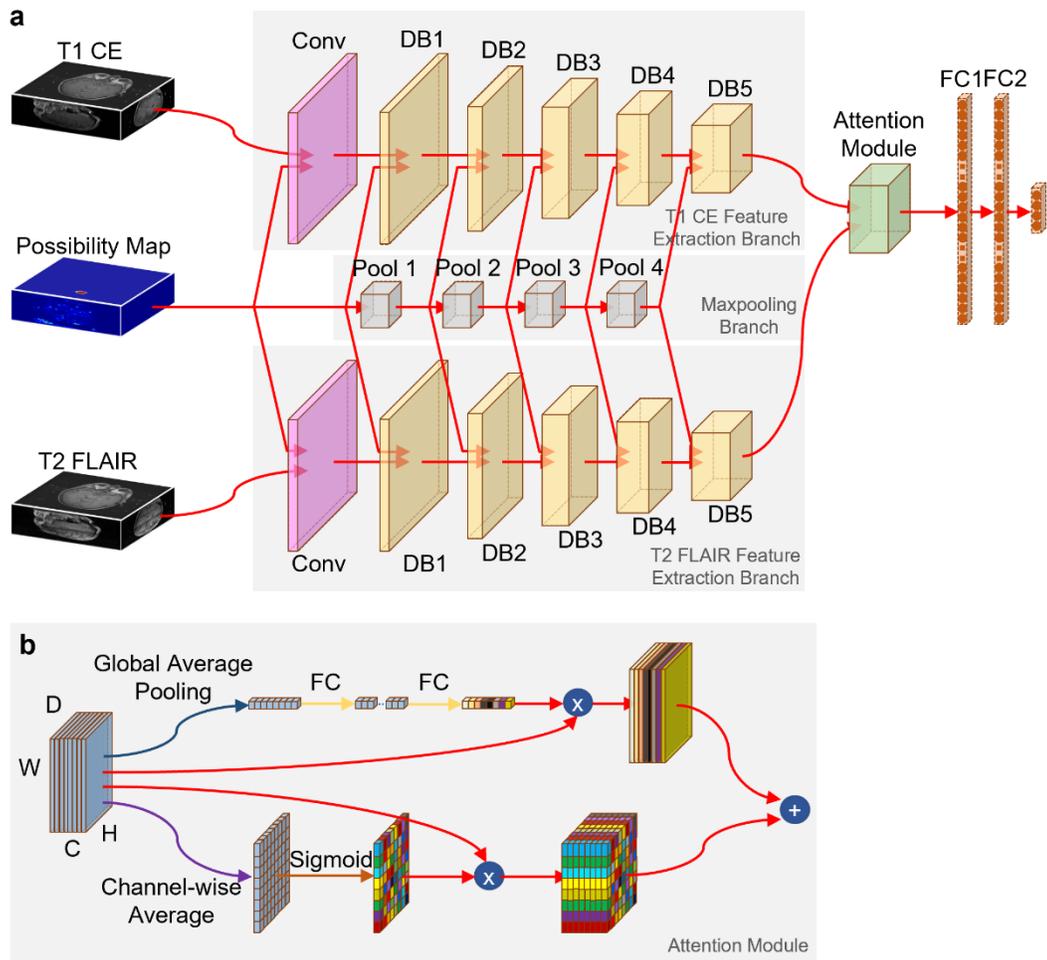

**Fig. 6. Proposed brain metastases classification network.** (a) The overall structure of the proposed classification network. (b) The structure of the attention module.

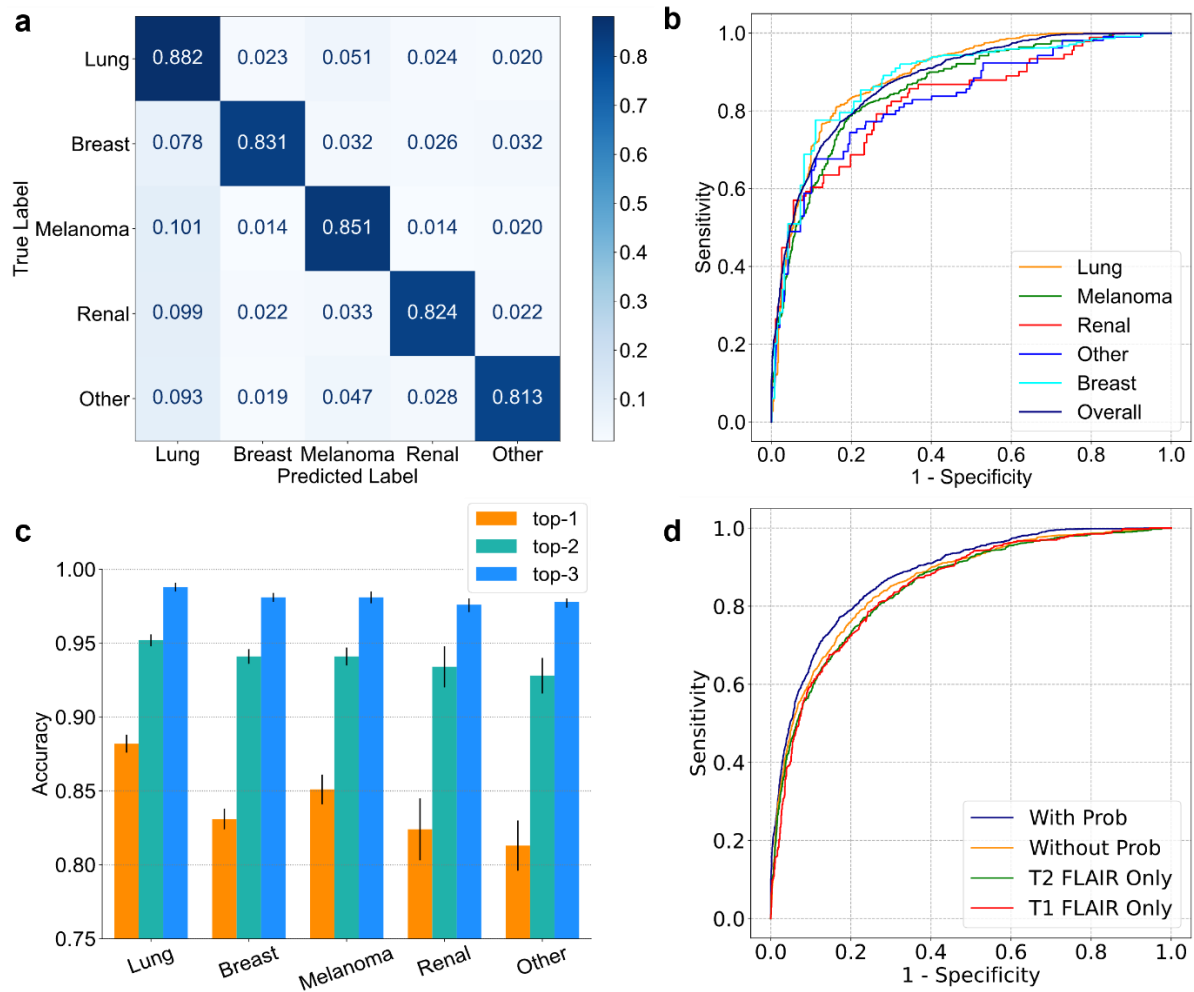

**Fig. 7. Five-class classification results.** (a) The confusion matrix of five-class classification based on 10-fold cross-validation. (b) Comparison of ROC curves from the five-class classification. (c) Top-k model accuracies for tumor origin classification for $k \in \{1, 2, 3\}$. (d) Comparison of ROC curves in the ablation study ("With Prob" shows the overall result, "Without Prob" shows the result without adding probability maps; "T2 FLAIR Only" shows the result only using T2 FLAIR images for the classification, and "T1 CE Only" shows the result only using T1 CE images for the classification).

**Table 1. Statistics of the 1,582 cases collected for this study.**

|  | With tumor contours | | Without tumor contours | | Total |
|---|---|---|---|---|---|
|  | T1 CE | T2 FLAIR | T1 CE | T2 FLAIR |  |
| **Lung** | 83 | 93 | 195 | 511 | 882 |
| **Breast** | 0 | 0 | 152 | 2 | 154 |
| **Melanoma** | 45 | 40 | 76 | 187 | 348 |
| **Renal** | 2 | 3 | 19 | 67 | 91 |
| **Other** | 8 | 9 | 23 | 67 | 107 |
| **Total** | 138 | 145 | 465 | 834 | 1582 |

**Table 2. Age and gender statistics of the 1,399 patients used in this study.**

|  | With tumor contours | Without tumor contours |
|---|---|---|
| **Ages** | 61.8 ± 11.2 | 62.9 ± 11.9 |
| **Gender** | M: 83  F: 65 | M: 739  F: 512 |

**Table 3. Quantitative comparison between the proposed segmentation network and three competing brain tumor segmentation networks.**

|  | Dice | |
|---|---|---|
|  | T1 CE | T2 FLAIR |
| **U-Net** | 0.866 ± 0.296 | 0.869 ± 0.287 |
| **AttU-Net** | 0.864 ± 0.312 | 0.878 ± 0.295 |
| **U-Net++** | 0.857 ± 0.325 | 0.880 ± 0.306 |
| **Proposed** | **0.878 ± 0.297** | **0.895 ± 0.282** |

**Table 4. Quantitative comparison of the 10-fold cross-validation results on five-class classification.**

|          | AUC   | 95% Confidence Interval |
|----------|-------|-------------------------|
| Lung     | 0.889 | 0.883 - 0.895           |
| Breast   | 0.873 | 0.860 - 0.886           |
| Melanoma | 0.852 | 0.842 - 0.862           |
| Renal    | 0.830 | 0.809 - 0.851           |
| Other    | 0.822 | 0.805 - 0.839           |
| Overall  | 0.878 | 0.873 - 0.883           |

**Table 5. Quantitative comparison of ablation study results on five-class classification.**

|               | AUC   | 95% Confidence Interval |
|---------------|-------|-------------------------|
| With Prob     | 0.878 | 0.873 - 0.883           |
| Without Prob  | 0.862 | 0.856 - 0.868           |
| T2 FLAIR Only | 0.851 | 0.844 - 0.858           |
| T1 CE Only    | 0.845 | 0.838 - 0.852           |

**Table 6. Quantitative comparison of the 10-fold cross-validation results on binary classification.**

|          | Breast | Melanoma | Other |
|----------|--------|----------|-------|
| Lung     | 0.959  | 0.941    | 0.938 |
| Breast   | -      | 0.945    | 0.942 |
| Melanoma | -      | -        | 0.937 |